\documentclass{aa}  
\bibpunct{(}{)}{;}{a}{}{,} 

\usepackage{graphicx}
\usepackage{txfonts}

\def\kms{km$\,$s\ensuremath{^{-1}}}

\def\Hb{${\rm H}{\beta}$}

\def\Mgb{{\rm Mg}$b$}
\def\Mg2{{\rm Mg}$_2$}
\def\Fe{$\langle {\rm Fe}\rangle$}
\def\MgFe{[${\rm MgFe}$]$'$}
\def\aFe{[$\alpha/{\rm Fe}$]~}

\def\ZH{[$Z/{\rm H}$]~}

\begin{document}

\title{The difference in age of the two counter-rotating stellar disks of
  the\\ spiral galaxy NGC~4138 \thanks{Based on observation carried out
    at the Galileo 1.22 m telescope at Padua University.}
  \thanks{Tables 1-3 are available in electronic form at
    http://www.aanda.org.}  }


\author{A. Pizzella\inst{1,2} 
     \and 
      L. Morelli\inst{1,2}
      \and
      E. M. Corsini\inst{1,2}
      \and
      E. Dalla Bont\`a\inst{1,2}
      \and
      L. Coccato\inst{3}
      \and
      G. Sanjana\inst{4}
      }

\institute{Dipartimento di Fisica e Astronomia `G. Galilei',
  Universit\`a di Padova, vicolo dell'Osservatorio 3, I-35122 Padova,
  Italy\\ 
\email{alessandro.pizzella@unipd.it}
\and INAF-Osservatorio Astronomico di Padova, 
  vicolo dell'Osservatorio 5, I-35122 Padova, Italy.
\and European Southern Observatory, 
  Karl-Schwarzschild-Stra$\beta$e 2, D-85748 Garching bei 
  M\"unchen, Germany
\and Department of Physics, Imperial College London, 
  South Kensington Campus, London SW7 2AZ, UK
}
\date{Received ...; accepted ...}

\abstract
%
{Galaxies accrete material from the environment through acquisitions
  and mergers. These processes contribute to the galaxy assembly and
  leave their fingerprints on the galactic morphology, internal
  kinematics of gas and stars, and stellar populations.}
%
%
{The Sa spiral NGC~4138 is known to host two counter-rotating stellar
  disks, with the ionized gas co-rotating with one of them. We measured
  the kinematics and properties of the two counter-rotating stellar
  populations to constrain their formation scenario.}
%
%
{A spectroscopic decomposition of the observed major-axis spectrum was
  performed to disentangle the relative contribution of the two
  counter-rotating stellar and one ionized-gas components. The
  line-strength indices of the two counter-rotating stellar
  components were measured and modeled with single stellar population
  models that account for the $\alpha/$Fe overabundance.}
%
%
{The counter-rotating stellar population is younger,
    marginally more metal poor, and more $\alpha$-enhanced than the
    main stellar component. The younger stellar component is also
  associated with a star-forming ring.}
%
%
{The different properties of the counter-rotating stellar components
  of NGC~4138 rule out the idea that they formed because of bar dissolution. Our
  findings support the results of numerical simulations in which the
  counter-rotating component assembled from gas accreted on retrograde
  orbits from the environment or from the retrograde merging with a
  gas-rich dwarf galaxy.}


\titlerunning{Counter-rotating stellar populations of NGC~4138}

\authorrunning{A. Pizzella et al.}

\maketitle

%

\section{Introduction}
\label{sec:introduction}

Counter-rotating galaxies host two components rotating in opposite
directions with respect to each other \citep{Bertola1999}.
Large-scale counter-rotating disks of stars and/or gas have been
detected in several lenticular and spiral galaxies, and different
mechanisms have been proposed to explain their formation
\citep{Corsini2014}. They are expected to leave different signatures
in the properties of the counter-rotating stellar populations. In
particular, their age difference may be used to discriminate between
competing scenarios for the origin of counter-rotation. Gas
accretion followed by star formation always predicts a younger age for
the counter-rotating component \citep{Pizzella2004, Vergani2007, Algorry2014},
whereas the counter-rotating component formed by the retrograde
capture of stars through minor or major mergers may be either younger
or older with respect to the pre-existing stellar disk
\citep{Thakar1997, Crocker2009}. The external origin also allows
the two counter-rotating components to have different metallicities and
$\alpha$-enhancements.  In contrast, the formation of large-scale
counter-rotating stellar disks due to bar dissolution predicts the
same mass, chemical composition, and age for both the prograde and
retrograde components \citep{Evans1994}.

We developed a spectroscopic decomposition that separates the relative
contribution of the counter-rotating stellar components to the
observed galaxy spectrum in order to constrain the properties of the
counter-rotating stellar populations \citep{Coccato2011}. We
successfully applied it to the counter-rotating stellar disks of
NGC~3593, NGC~4550, and NGC~5719 \citep{Coccato2011, Coccato2013}. In
all of them, the counter-rotating stellar disk rotates in the same
direction as the ionized gas, and it is less massive, younger, more
metal poor, and more $\alpha$-enhanced than the main stellar disk.
These findings rule out an internal origin of the secondary stellar
component and favor a scenario where it formed from gas accreted on
retrograde orbits from the environment, which fueled rapid star
formation. The merging scenario cannot be ruled out, because it also
allows the counter-rotating component to be the younger one in half of
the cases. The available statistics are not sufficient to draw
significant conclusions and therefore a larger sample is required to
address this issue.
  
In this paper our spectroscopic technique is applied to derive
the properties of the counter-rotating stellar populations of the Sa
galaxy NGC~4138. The structure, kinematics, and dynamics of this
nearby and relatively isolated galaxy were studied in detail by
\citet{Jore1996} and \citet{Afanasiev2002}. Optical images show no
major peculiarities except for a narrow dust lane in the galactic
disk. Radio observations show an extended and warped disk of neutral
hydrogen which rotates in the same direction as the ionized gas and
counter-rotates with respect to the main stellar component
\citep{Jore1996}.

This paper is organized as follows: the spectroscopic data of NGC~4138
are presented in Sect.~\ref{sec:data}. The kinematic and stellar
population properties of the galaxy are measured and discussed in
Sect.~\ref{sec:analysis}. The conclusions about the formation scenario
of the counter-rotating components of NGC~4138 are given in
Sect.~\ref{sec:conclusions}.

\section{Observations and data reduction}
\label{sec:data}

The spectroscopic observations of NGC~4138 were carried out at the
Asiago Astrophysical Observatory of Padua University with the
1.22 m Galileo telescope on December 12, 2012.
The telescope was equipped with the Cassegrain Boller \& Chivens
spectrograph. The grating with 1200 grooves mm$^{-1}$ was used in the
first order in combination with a $3\farcs0\, \times\, 7\farcm75$ slit
and an Andor iDus DU440 CCD, which has $2048\, \times\, 512$ pixels of
$26\, \times\, 26$ $\rm \mu m^2$. The gain and readout noise were 0.97
$e^-$ count$^{-1}$ and 3.4 $e^-$ (rms), respectively.
The spectral range between about $4500$ \AA\ and $5700$ \AA\ was
covered with a reciprocal dispersion of $0.60$ \AA\ pixel$^{-1}$. The
spatial scale was $1\farcs0$ pixel$^{-1}$.
The instrumental resolution was $2.47\pm0.18$ \AA\ ({\em FWHM\/}), and
it was derived from the mean of the Gaussian {\em FWHM\/} that were
measured for a dozen unblended sky emission lines distributed over the
whole spectral range of a wavelength-calibrated spectrum.  It
corresponds to $\sigma_{\rm instr} = 63\pm4$ \kms\ at 5000 \AA .

\begin{figure}[ht!]
 \centering
 \includegraphics[width=9cm]{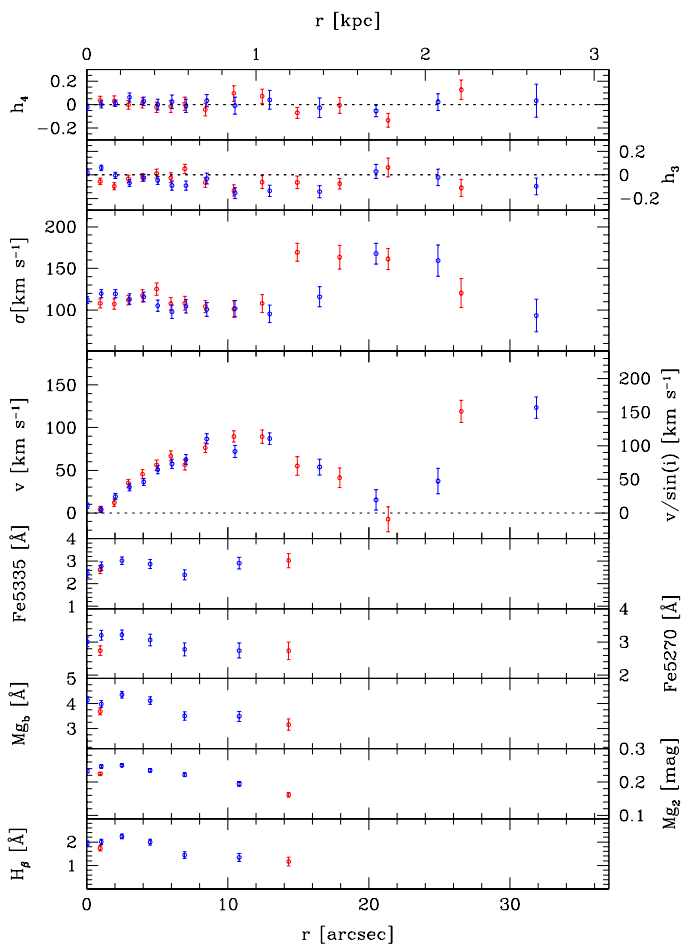}
 \caption{Stellar kinematics and line-strength indices as measured
   along the major axis of NGC~4138. The radial profiles of the
   line-of-sight fourth- and third-order coefficient of the
   Gauss-Hermite decomposition of the LOSVD ($h_4$ and $h_3$);
   velocity dispersion ($\sigma$); velocity (v) after the
   subtraction of systemic velocity; and the line-strength indices
   Fe$_{5335}$, Fe$_{5270}$, \Mgb, \Mg2, and \Hb\ are plotted ({\em
     from top to bottom}). The red and blue dots indicate the receding
   (NW) and approaching (SE) side of the galaxy, respectively.}
\label{fig:kinematics}
\end{figure}

The galaxy was observed along the major axis ($\rm PA = 150\degr$).  At the
beginning of each exposure, the galaxy was centered on the slit using
the guiding camera. Seven exposures of 1800 s each ensured 3.5 hours
of effective integration. HR8634 was observed as spectro-photometric
standard star to calibrate the flux of the galaxy spectra before
line-strength indices were measured. Spectra of the comparison arc
lamp were taken before and after object exposures.
The value of the seeing {\em FWHM} during the observing nights, as
measured on the guiding star, was about $4\arcsec$.

All the spectra were bias subtracted, flat-field corrected, cleaned of
cosmic rays, corrected for bad columns, and wavelength and flux
calibrated using IRAF\footnote{Image Reduction and Analysis Facility
  (IRAF) is distributed by the National Optical Astronomy
  Observatories, which are operated by the Association of Universities
  for Research in Astronomy, Inc., under cooperative agreement with
  the National Science Foundation.}. Each spectrum was rebinned using
the wavelength solution obtained from the corresponding arc-lamp
spectrum obtained as the average of the arc-lamp observed immediately
before and after the science exposure.
All the galaxy and stellar spectra were corrected for CCD
misalignment. The sky contribution was determined by interpolating
along the two edges of the slit, where the stellar light of the galaxy was
negligible, and then subtracted.  The different galaxy spectra were
co-added into a single major-axis spectrum using the center of the
stellar continuum as a reference.

\section{Analysis and results}
\label{sec:analysis}

The galaxy spectrum was first analyzed without disentangling the two
counter-rotating components; afterwards the stellar population
properties of both components were derived by performing a
spectroscopic decomposition.

\subsection{Single-component analysis and results}
\label{sec:onec}

The stellar kinematics (Fig.~\ref{fig:kinematics}) were measured
following \citet{Pizzella2013} using the Penalized Pixel Fitting 
\citep[pPXF,][]{Cappellari2004} and Gas and Absorption Line
Fitting \citep[GANDALF; ][]{Sarzi2006} IDL\footnote{The Interactive
  Data Language (IDL) is distributed by ITT Visual Information
  Solutions.} codes.
The galaxy spectrum was rebinned along the dispersion direction to a
logarithmic scale and along the spatial direction to obtain a
signal-to-noise ratio $S/N \geq 15$ per resolution element when
measuring the kinematic parameter.
At each radius, a linear combination of single stellar population models from
\citet{maraston2011} was convolved with the
line-of-sight velocity distribution (LOSVD) assumed to be a Gaussian
plus third- and fourth-order Gauss-Hermite polynomials and fitted to
the galaxy spectrum by $\chi^2$ minimization in pixel space. 

The stellar kinematics are given in Table~1 and extend out to
$|r|\simeq30\arcsec$ on both sides of the nucleus. The velocity curve
is symmetric around the center and the deprojected velocity rises to
$\simeq100$ \kms\ at $|r|\simeq12\arcsec$. Farther out it decreases to
$\sim0$ \kms\ at $|r|\simeq21\arcsec$.  For larger radii the velocity
increases again to $\simeq100$ \kms.  The velocity dispersion is
$\simeq120$ \kms\ in the center and slightly decreases outwards. It
increases again for $|r|>14$\arcsec\ peaking at $\simeq170$ \kms\ at
$|r|\simeq21$\arcsec.  This behavior of both the velocity and velocity
dispersion radial profiles is the signature of the presence of
a counter-rotating stellar component \citep{Bertola1996,
  Vergani2007}. The relative contribution of the counter-rotating
component to the galaxy surface brightness is maximum at
$|r|\simeq21\arcsec$ \citep{Jore1996}. In this radial range a star-forming ring of gas
\citep{Pogge1987, Jore1996} and stars \citep{Jore1996, Afanasiev2002}
was detected. The main stellar component dominates the light
distribution of the inner regions ($|r|<12\arcsec$).

The Mg, Fe, and \Hb\ line-strength indices (Fig.~\ref{fig:kinematics})
were measured from the absorption features of the galaxy spectrum
following \citet{Morelli2012}. The galaxy spectrum was rebinned in the
radial direction to achieve a $S/N \geq 25$ per resolution element.
The spectral resolution of the galaxy spectrum was degraded through a
Gaussian convolution to match the Lick/IDS resolution before measuring
the line-strength indices. 

The values of the line-strength indices
measured for a sample of templates were compared to those obtained by
\citet{Worthey1994} to calibrate our measurements to the Lick/IDS
system. The offsets were neglected, being smaller than the mean error
of the differences. The average iron index \Fe\ \citep{Gorgas1990} and
combined magnesium-iron index \MgFe\ \citep{Thomas2003} were
calculated to be used in combination with the \Mgb\ and \Hb\ indices
to constrain the stellar population properties.

A foreground star is observed at $r\simeq4\arcsec$ NW from the center
of NGC~4138 roughly along its major axis \citep{Afanasiev2002}. The
star spectrum affects the line-strength measurements since it changes
the observed continuum level, lowering the strength of the galaxy
absorption features. The large difference between the redshifts of the
star and galaxy allowed the correct measurement of the galaxy
kinematics.
Only the line-strength measurements not affected by the foreground
star are shown in Fig.~\ref{fig:kinematics}. They are listed in
Table 2 and are found to be in agreement with the measurements by
\citet{Afanasiev2002}.

The central line-strength indices fit within the scatter
the \Mg2-$\sigma$, \Fe-$\sigma$, and \Hb-$\sigma$ relations for 
early-type spiral galaxies by \citet{Morelli2012}.

\begin{figure}[ht!]
\centering
\includegraphics[width=9cm]{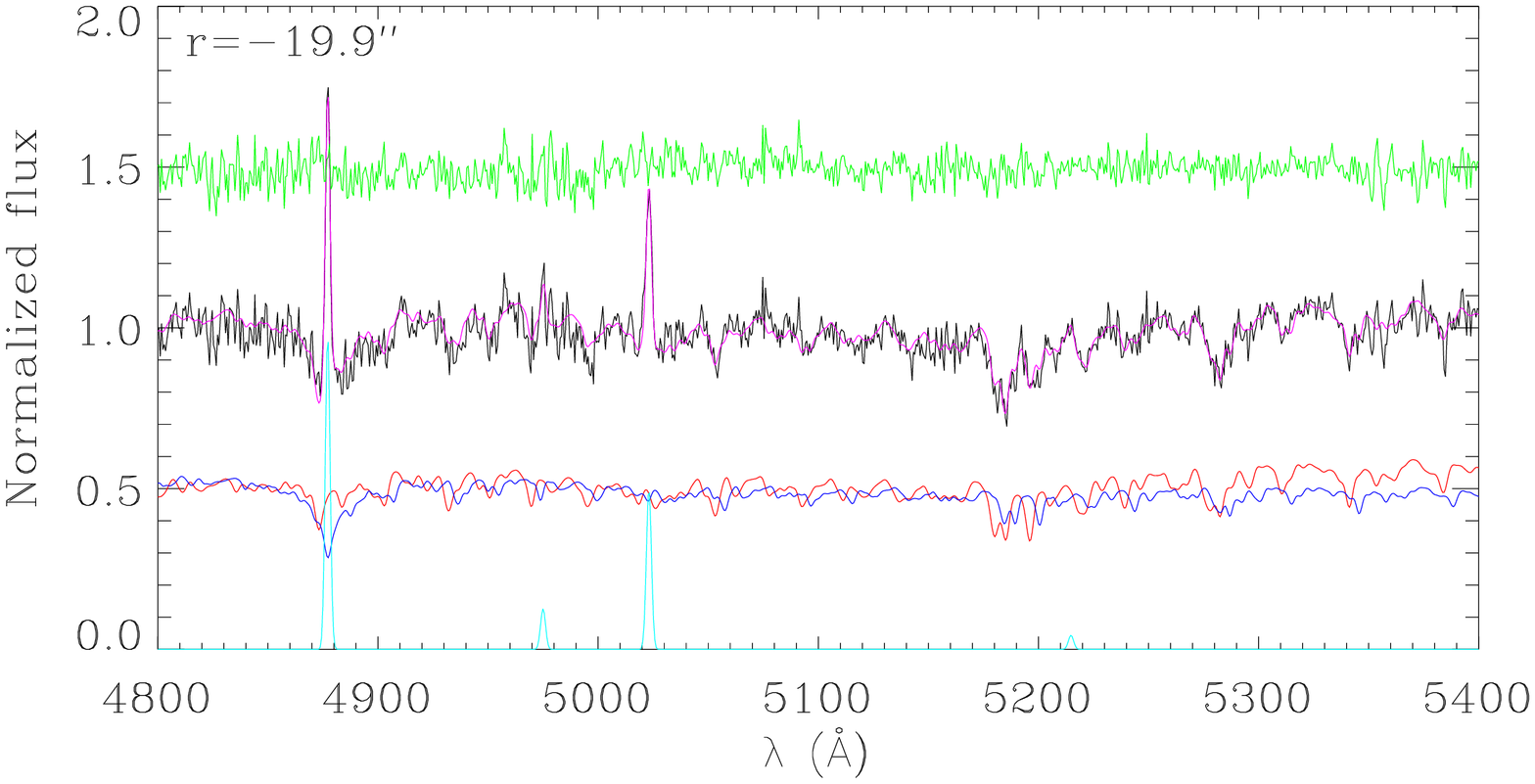}
\includegraphics[width=9cm]{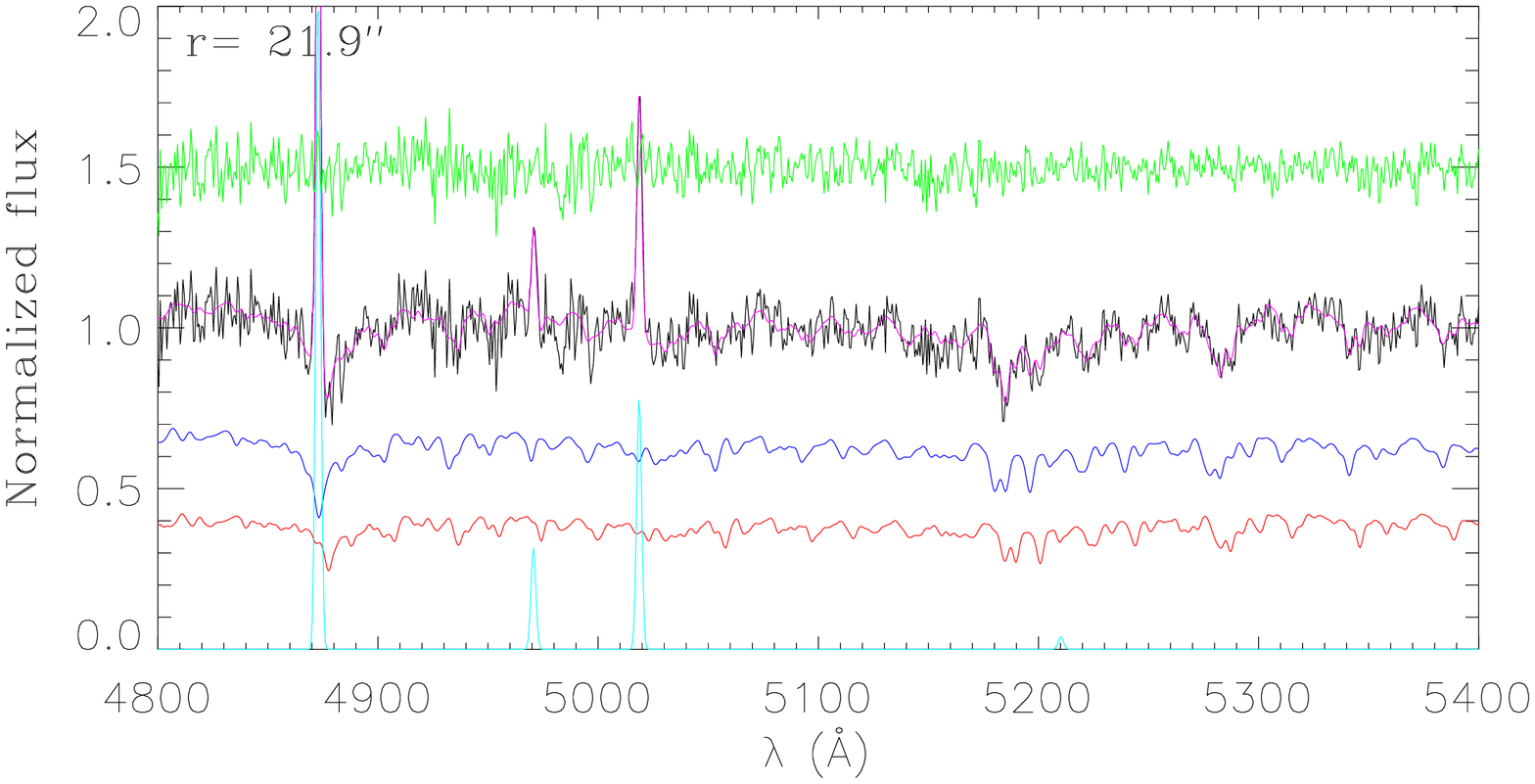}
\caption{Fit of the averaged spectrum (black line) in the spatial bin
  at $-19.9\arcsec$ ({\em top panel}) and $+21.9\arcsec$ ({\em bottom
    panel}) from the galaxy center. The best-fitting model (magenta
  line) is the sum of the spectra of the ionized-gas component (cyan
  line) and the two stellar components (blue and red lines). The
  normalized flux of the fit residual (green line) has a false zero
  point for viewing convenience. The double peaked absorption lines
  are evident in the averaged spectra.}
\label{fig:fit}
\end{figure}

\subsection{Two-component analysis and results}
\label{sec:twoc}

A high spectral $S/N$ is needed to perform a spectral decomposition.
Therefore, we averaged the galaxy spectrum along the spatial direction
between $r=-25\arcsec$ and $r=-16\arcsec$, and between $r=17\arcsec$
and $r=28\arcsec$ (Fig.~\ref{fig:fit}). The resulting spectra sample
the two radial ranges where the contribution of the counter-rotating
component is at its maximum.

The two spectra were analyzed with the novel implementation pPXF developed
in \citet{Coccato2011}. It fit\-ted to the galaxy spectrum two synthetic
templates (one for each stellar component) obtained as a linear
combination of models by \citet{maraston2011} by convolving them with two
Gaussian LOSVDs with different kinematics. The relative contribution
of the counter-rotating components to the total spectrum was in terms
of light. Gaussian functions were added to the convolved synthetic
templates to account for ionized-gas emission lines. The spectroscopic
decomposition returned the spectra of two best-fit synthetic stellar
templates and ionized-gas emissions, along with the best-fitting
parameters of luminosity fraction and line-of-sight velocity and
velocity dispersion.
In both spatial bins the averaged spectrum is the superposition of two
stellar components which are rotating in opposite directions
(Fig.~\ref{fig:fit}). One component is co-rotating with the ionized
gas and main body of the galaxy, while the other one is
counter-rotating. In particular we measured a mean velocity of
  $-116\pm 5$ \kms\ at $r=-19\farcs9$ and $153\pm 8$ \kms\ at
  $r=+21\farcs9$ for the main component and a mean velocity of $135\pm
  10$ \kms\ at $r=-19\farcs9$ and $-117\pm 8$ \kms\ at $r=+21\farcs9$
  for the secondary component. The ionized gas has about the same
  velocity as the counter-rotating component
  (Fig.~\ref{fig:fit}).

Both components contribute about $50\%$ of the
galaxy luminosity in the binned radial range. The line-of-sight
velocities and velocity dispersions measured for the ionized gas and
the two stellar components are consistent with \citet{Jore1996}.

\begin{figure*}[ht!]
  \centering
  \includegraphics[angle=90,width=0.48\textwidth]{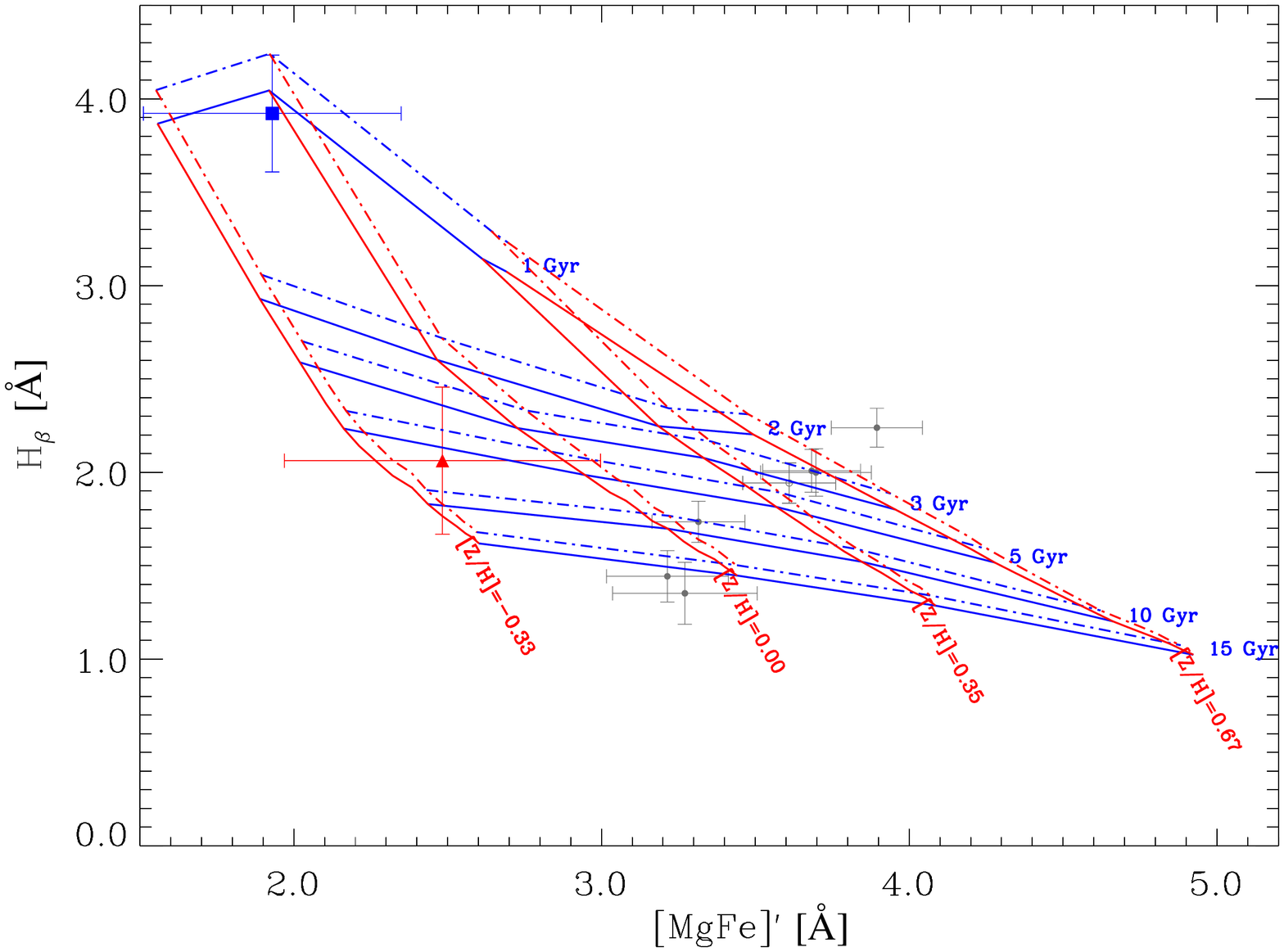}
  \includegraphics[angle=90,width=0.48\textwidth]{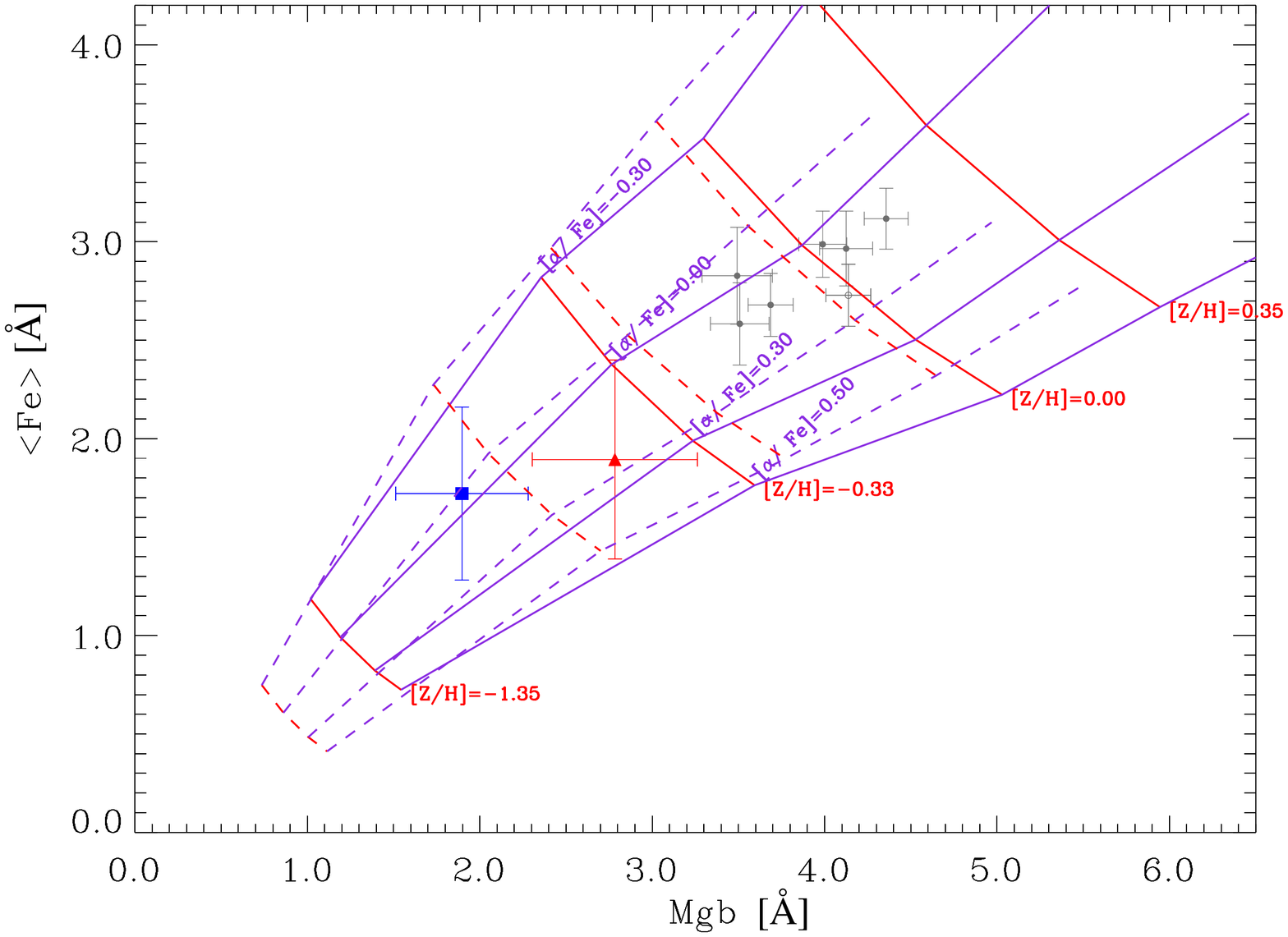}
  \caption{\Hb\ and \MgFe\ indices ({\em left panel}) and \Mgb\ and \Fe\ 
    indices ({\em right} panel) measured along the major axis of
    NGC~4138. The red triangles and blue squares correspond to the
    values measured at $|r|\simeq21\arcsec$ for the co-rotating and
    counter-rotating components, respectively. The gray filled circles
    are the line strengths measured in the inner $12\arcsec$ where
    only the main component is present. The center of the galaxy is
    indicated by an open gray circle. The lines indicate the models by
    \cite{Thomas2003}. In the left panel the age-metallicity grids are
    plotted with two different $\alpha$/Fe enhancements:
    \aFe$\,=\,0.0$ dex (continuous lines) and \aFe$\,=\,0.5$ dex
    (dashed lines). In the right panel the \aFe\ ratio-metallicity
    grids are plotted with two different ages: 2 Gyr (continuous
    lines) and 12 Gyr (dashed lines).} 
   \label{fig:ssp}
\end{figure*}

The line-strength indices of the two counter-rotating components were
extracted from the two best-fit synthetic templates. The values
measured for each component on the two galaxy sides were averaged and
are given in Table 3.
In Fig.~\ref{fig:ssp} these measurements are compared to the
line-strength indices predicted for a single stellar population as
a function of the age, metallicity, and \aFe ratio by
\citet{Thomas2003}. The line-strength indices derived in the inner
$12\arcsec$ where the main stellar component is dominant are also
plotted in Fig.~\ref{fig:ssp}. They are representative of the main
body of the galaxy.

The two counter-rotating stellar components are characterized by
 marginally different chemical properties. The co-rotating stellar component is
characterized by higher values of \MgFe, \Mgb\, and \Fe , and a lower
\Hb\ value with respect to the counter-rotating component. This
translates into different properties of their stellar populations,
which were derived by a linear interpolation of measured line
strengths between the model predictions using an iterative procedure
as in \citet{Morelli2008}.
The co-rotating stellar component is older ($6.6\pm3.6$ Gyr) with
sub-solar metallicity (\ZH$\,=-0.24\pm0.46$ dex) and super-solar
enhancement (\aFe$\,=0.24\pm0.19$ dex), whereas the counter-rotating
stellar population component is younger ($1.1\pm0.3$ Gyr) with solar
metallicity (\ZH$\,= -0.04\pm0.27$ dex) and solar enhancement
(\aFe$\,= 0.08\pm0.21$ dex). The age of the co-rotating component
matches the age range ($2-15$ Gyr) measured for the main body of the
galaxy.

\section{Discussion and conclusions}
\label{sec:conclusions}

The Sa spiral NGC~4138 hosts two cospatial counter-rotating stellar
disks, one of which is co-rotating with the gaseous disk
\citep{Jore1996, Afanasiev2002}.

We have decomposed an intermediate-resolution spectrum measured along
the major axis of NGC~4138 to derive the properties of the
counter-rotating stellar populations at $|r|\simeq21\arcsec$. In this
radial range the counter-rotating components give about the same
contribution to the surface brightness of the galaxy.
We have found that the counter-rotating stars are on average younger,
more metal poor, and more $\alpha$-enhanced than those in the main
component. These features were also observed in the few other galaxies
where the stellar populations of counter-rotating stellar disks of
comparable sizes were successfully disentangled (NGC~3593,
\citealt{Coccato2013}; NGC~4550, \citealt{Coccato2013} and
\citealt{Johnston2013}; NGC~5719, \citealt{Coccato2011}).
The difference in age between the two counter-rotating stellar
components is particularly strong in NGC~5179, which shows a
spectacular on-going interaction with its nearby companion
NGC~5713. The two galaxies are connected by a tidal bridge of neutral
hydrogen which is feeding the counter-rotating gaseous and stellar
components \citep{Vergani2007}. On the contrary, NGC~3593 and NGC~4550
are both quite isolated and undisturbed galaxies. The age difference
between the two stellar populations is less pronounced in NGC~4550,
whereas NGC~3593 represents an intermediate case.

The counter-rotating stellar populations of NGC~4138 have almost the
same age as those detected in NGC~3593. It is worth noticing that the
two galaxies are characterized by similar morphological and structural
properties. Indeed, they are very early-type spirals with smooth arms
defined by dust lanes \citep{Sandage1994}. Moreover, both NGC~4138
\citep{Pogge1987, Jore1996, Afanasiev2002} and NGC~3593
\citep{Corsini1998, Garcia-Burillo2000, Coccato2013} show a
star-forming ring of gas and stars in the region where the younger
stellar component is detected.

The suppression of arms in counter-rotating spirals has been recently
recognized in high-resolution N-body simulations of multi-armed spiral
features triggered through swing amplification by density
inhomogeneities orbiting the disk \citep{DOnghia2013}. However, a
survey of a sample of early-type spirals selected to have the spiral
pattern traced by dust lanes revealed the presence of kinematically
decoupled gas components but no new case of counter-rotation
\citep{Corsini2003}.

The difference in age of the counter-rotating components of NGC~4138 rules
out the possibility that their formation was driven by internal
mechanisms and supports a process involving acquisition of material
from the environment and merging events.
Using numerical simulations, \citet{Thakar1997} modeled the formation
process of NGC~4138 by investigating the continuous retrograde infall
of gas and the retrograde merger with a gas-rich dwarf galaxy. Both
processes are successful in producing a counter-rotating disk of the
observed mass and size without heating up the primary disk
significantly. The timescale for the case of continuous infall ($5$
Gyr) was found to be somewhat longer than for a dwarf merger ($4$ Gyr),
but it is still consistent with the age difference actually measured
for the two counter-rotating populations. According to the simulations
by \citet{Thakar1997}, the ring of NGC~4138 was created by
counter-rotating gas clouds colliding with co-rotating gas already
present in the disk and forming stars in the process. This scenario is
supported by our findings that the star-forming ring is associated with
the younger stellar component.

A few decades after they were discovered, counter-rotating
galaxies still represent a challenging subject. Deriving the
properties of their stellar populations with spectral decomposition
makes it possible to test the predictions of the different scenarios
suggested to explain their formation.

\begin{acknowledgements}
This work was supported by Padua University through grants
60A02-5052/11, 60A02-4807/12, 60A02-5857/13, and
CPDA133894. L.M. received financial support from Padua University
grant CPS0204. G.S. acknowledges Padua University for their hospitality 
while this paper was in progress.
\end{acknowledgements}

\bibliographystyle{aa} 

\newpage
\onecolumn
\begin{center}
{\bf On-line tables}
\end{center}

\setcounter{table}{0}

\begin{table}[ht!]
\caption{Stellar kinematics along the major axis of NGC~4138}
\begin{tabular}{rrrrr}
\hline
\noalign{\smallskip}
\hline
\noalign{\smallskip}
\multicolumn{1}{c}{$r$} &
\multicolumn{1}{c}{$V$} &
\multicolumn{1}{c}{$\sigma$} &
\multicolumn{1}{c}{$h_3$} &
\multicolumn{1}{c}{$h_4$} \\
\noalign{\smallskip}
\multicolumn{1}{c}{[\arcsec]} &
\multicolumn{1}{c}{[\kms]} &
\multicolumn{1}{c}{[\kms]} &
\multicolumn{1}{c}{} &
\multicolumn{1}{c}{} \\
\noalign{\smallskip}
\hline
\noalign{\smallskip}
$-31.9$ & $-124\pm12$ & $ 93\pm20$ & $ 0.097\pm0.071$ & $ 0.034\pm0.141$\\ 
$-24.9$ & $ -37\pm15$ & $159\pm19$ & $ 0.020\pm0.070$ & $ 0.022\pm0.072$\\ 
$-20.5$ & $ -15\pm12$ & $168\pm13$ & $-0.030\pm0.061$ & $-0.054\pm0.050$\\ 
$-16.5$ & $ -54\pm 9$ & $116\pm12$ & $ 0.142\pm0.051$ & $-0.026\pm0.084$\\ 
$-13.0$ & $ -87\pm 7$ & $ 95\pm10$ & $ 0.136\pm0.048$ & $ 0.040\pm0.080$\\ 
$-10.5$ & $ -72\pm 7$ & $102\pm10$ & $ 0.154\pm0.046$ & $-0.009\pm0.073$\\ 
$ -8.5$ & $ -87\pm 6$ & $101\pm 8$ & $ 0.031\pm0.047$ & $ 0.033\pm0.052$\\ 
$ -7.0$ & $ -63\pm 6$ & $104\pm 8$ & $ 0.091\pm0.040$ & $-0.012\pm0.057$\\ 
$ -6.0$ & $ -58\pm 5$ & $ 98\pm 8$ & $ 0.090\pm0.039$ & $ 0.025\pm0.058$\\ 
$ -5.0$ & $ -51\pm 5$ & $105\pm 7$ & $ 0.045\pm0.034$ & $ 0.001\pm0.047$\\ 
$ -4.0$ & $ -36\pm 4$ & $116\pm 6$ & $ 0.023\pm0.030$ & $ 0.028\pm0.035$\\ 
$ -3.0$ & $ -30\pm 4$ & $113\pm 6$ & $ 0.068\pm0.029$ & $ 0.062\pm0.039$\\ 
$ -2.0$ & $ -19\pm 4$ & $120\pm 5$ & $ 0.004\pm0.026$ & $ 0.013\pm0.028$\\ 
$ -1.0$ & $  -4\pm 3$ & $120\pm 5$ & $-0.061\pm0.024$ & $ 0.002\pm0.030$\\ 
$ -0.0$ & $  -8\pm 3$ & $112\pm 4$ & $-0.020\pm0.025$ & $-0.023\pm0.026$\\ 
$  1.0$ & $   4\pm 4$ & $108\pm 6$ & $-0.055\pm0.027$ & $ 0.035\pm0.036$\\ 
$  2.0$ & $  12\pm 4$ & $107\pm 6$ & $-0.097\pm0.031$ & $ 0.031\pm0.043$\\ 
$  3.0$ & $  35\pm 4$ & $112\pm 6$ & $-0.032\pm0.031$ & $-0.001\pm0.037$\\ 
$  4.0$ & $  45\pm 6$ & $118\pm 7$ & $-0.022\pm0.034$ & $ 0.016\pm0.046$\\ 
$  5.0$ & $  56\pm 6$ & $125\pm 8$ & $ 0.012\pm0.038$ & $-0.022\pm0.045$\\ 
$  6.0$ & $  67\pm 6$ & $108\pm 8$ & $-0.024\pm0.042$ & $-0.015\pm0.052$\\ 
$  7.0$ & $  56\pm 6$ & $108\pm 8$ & $ 0.052\pm0.039$ & $ 0.007\pm0.057$\\ 
$  8.4$ & $  76\pm 6$ & $104\pm 7$ & $-0.064\pm0.041$ & $-0.042\pm0.053$\\ 
$ 10.4$ & $  90\pm 7$ & $101\pm10$ & $-0.133\pm0.048$ & $ 0.096\pm0.066$\\ 
$ 12.4$ & $  90\pm 8$ & $108\pm11$ & $-0.062\pm0.055$ & $ 0.072\pm0.061$\\ 
$ 14.9$ & $  55\pm11$ & $169\pm11$ & $-0.063\pm0.054$ & $-0.070\pm0.047$\\ 
$ 17.9$ & $  41\pm12$ & $164\pm14$ & $-0.074\pm0.045$ & $-0.006\pm0.067$\\ 
$ 21.4$ & $  -7\pm15$ & $161\pm13$ & $ 0.064\pm0.080$ & $-0.133\pm0.058$\\ 
$ 26.6$ & $ 119\pm13$ & $120\pm17$ & $-0.109\pm0.073$ & $ 0.127\pm0.084$\\ 
\noalign{\smallskip}
\hline
\end{tabular}
\end{table}

\begin{table}[h!]
\caption{Line-strength indices along the major axis of NGC~4138}
\begin{tabular}{rrcccc}
\hline
\noalign{\smallskip}
\hline
\noalign{\smallskip}
\multicolumn{1}{c}{$r$} &
\multicolumn{1}{c}{\Hb} &
\multicolumn{1}{c}{\Mg2} &
\multicolumn{1}{c}{\Mgb} &
\multicolumn{1}{c}{Fe$_{5270}$} &
\multicolumn{1}{c}{Fe$_{5335}$}\\
\noalign{\smallskip}
\multicolumn{1}{c}{[\arcsec]} &
\multicolumn{1}{c}{[\AA]} &
\multicolumn{1}{c}{[mag]} &
\multicolumn{1}{c}{[\AA]} &
\multicolumn{1}{c}{[\AA]} &
\multicolumn{1}{c}{[\AA]} \\
\noalign{\smallskip}
\hline
\noalign{\smallskip} 

$-10.8$&$ 1.35\pm 0.17$&$ 0.20\pm 0.01$&$ 3.49\pm 0.20$&$ 2.74\pm 0.23$&$ 2.91\pm 0.27$\\ 
$ -6.9$&$ 1.44\pm 0.14$&$ 0.22\pm 0.01$&$ 3.51\pm 0.17$&$ 2.77\pm 0.19$&$ 2.39\pm 0.23$\\ 
$ -4.5$&$ 2.00\pm 0.13$&$ 0.24\pm 0.01$&$ 4.12\pm 0.15$&$ 3.06\pm 0.17$&$ 2.87\pm 0.21$\\ 
$ -2.5$&$ 2.24\pm 0.10$&$ 0.25\pm 0.01$&$ 4.36\pm 0.13$&$ 3.21\pm 0.14$&$ 3.02\pm 0.17$\\ 
$ -1.0$&$ 2.01\pm 0.12$&$ 0.25\pm 0.01$&$ 3.99\pm 0.14$&$ 3.20\pm 0.15$&$ 2.77\pm 0.19$\\ 
$ -0.0$&$ 1.94\pm 0.11$&$ 0.23\pm 0.01$&$ 4.14\pm 0.13$&$ 3.00\pm 0.14$&$ 2.45\pm 0.18$\\ 
$  1.0$&$ 1.73\pm 0.11$&$ 0.23\pm 0.01$&$ 3.69\pm 0.13$&$ 2.74\pm 0.14$&$ 2.62\pm 0.18$\\ 
$ 14.3$&$ 1.17\pm 0.18$&$ 0.16\pm 0.01$&$ 3.16\pm 0.22$&$ 2.73\pm 0.26$&$ 3.03\pm 0.31$\\  

\noalign{\smallskip} 
\hline                                                                          
\end{tabular}
\label{tab:lick1c}
\end{table}

\begin{table}[h!]
\caption{Line-strength indices at $|r|\simeq21\arcsec$ for the
  co-rotating (no. 1) and counter-rotating (no. 2) stellar components of
  NGC~4138}
\begin{tabular}{rccccc}
\hline
\noalign{\smallskip}
\hline
\noalign{\smallskip}
\multicolumn{1}{c}{No.} &
\multicolumn{1}{c}{\Hb} &
\multicolumn{1}{c}{\Mg2} &
\multicolumn{1}{c}{\Mgb} &
\multicolumn{1}{c}{Fe$_{5270}$} &
\multicolumn{1}{c}{Fe$_{5335}$}\\
\noalign{\smallskip}
\multicolumn{1}{c}{} &
\multicolumn{1}{c}{[\AA]} &
\multicolumn{1}{c}{[mag]} &
\multicolumn{1}{c}{[\AA]} &
\multicolumn{1}{c}{[\AA]} &
\multicolumn{1}{c}{[\AA]} \\
\noalign{\smallskip}
\hline
\noalign{\smallskip} 
1 & $2.06\pm0.39$ & $0.18\pm0.01$ & $2.78\pm0.48$ & $2.14\pm0.47$ & $1.65\pm0.54$\\ 
2 & $3.92\pm0.31$ & $0.10\pm0.01$ & $1.90\pm0.38$ & $1.85\pm0.39$ & $1.59\pm0.49$\\ 
\noalign{\smallskip}
\hline
\end{tabular}
\label{tab:lick2c}
\end{table}

\end{document}